\documentclass[prl,twocolumn,showpacs,noshowkeys,preprintnumbers,amsmath,amssymb]{revtex4}

\usepackage{graphicx}% Include figure files
\usepackage{dcolumn}% Align table columns on decimal point
\usepackage{bm}% bold math
\usepackage{hyperref}%
\usepackage{epsfig}
\begin{document}

\preprint{}

%\documentclass[prb,twocolumn,showpacs,noshowkeys,preprintnumbers,amsmath,amssymb]{revtex4}

%\documentclass[prb,twocolumn,showpacs,noshowkeys,preprintnumbers,amsmath,amssymb]{revtex4}
%\documentclass[aps,prb,showpacs,12pt]{revtex4}
%\documentclass[prb,preprint,showpacs,noshowkeys,preprintnumbers,amsmath,amssymb]{revtex4}

%\DeclareOption{a4paper}{

%\setleftmargin{19.1}
%\setrightmargin{19.1}

%\settopmargin{49pt}
%\setbottommargin{7pt}
	%\setallmargins{dimen}

%\setlength\paperheight {297mm}
%\setlength\paperwidth {210mm}

%\usepackage{graphicx}% Include figure files
%\usepackage{dcolumn}% Align table columns on decimal point
%\usepackage{bm}% bold math
%\usepackage{hyperref}%
%\nonstopmode
%\nofiles
%\bibliographystyle{apsrev}
%\usepackage{epsfig}

%\begin{document}

%\topmargin 75pt

%\preprint{}
%\title{Localization lengths of quasi-one dimensional disordered systems:
%Theory and numerical %analysis}
%
\title{Localization Length in Quasi One Dimensional Disordered System Revised}
\author{Vladimir Gasparian}
%\email{vgasparyan@csub.edu}
\affiliation{Department of Physics, California State University, Bakersfield, CA 93311, USA}
\author{Emilio Cuevas}
%\email{ecr@um.es}
%%%\homepage{http://bohr.fcu.um.es/miembros/ecr/}
\affiliation{Departamento de F{\'\i}sica, Universidad de Murcia, E-30071 Murcia, Spain}

%\date{\today}

\begin{abstract}
In the weak disordered regime we provide analytical expressions for the electron
localization lengths in quasi-one dimensional (Q1D) disordered quantum wire with
hard wall and periodic boundary conditions. They are exact up to order $W^2$
($W$ being the disorder strength) for an arbitrary number of channels. Detailed
numerical analysis of the Anderson localization, based on Kubo's formula for
conductivity, show excellent agreement with analytical calculations. We establish
relationship between various lengths in Q1D systems.
\end{abstract}
\pacs{72.15.Rn, 73.20.Fz,73.23.-b} 
\maketitle
\textit{Introduction}$.-$ A quasi-one dimensional (Q1D) geometry, as a model for
a disordered wire, is of great interest in condensed
matter theory. The electronic transport problem in weakly disordered Q1D systems can be solved
analytically within some approximations (see, e.g., \cite{markos} for details). The Dorokhov-Mello-Pereyra-Kumar (DMPK) equation \cite{1} and
random matrix theory for the transfer matrix \cite{2}
are the two successful approaches which are generally applied to describe the
behavior of conductance in a disordered wire. These two approaches give
very similar solutions for the probability distribution of conductance
in Q1D and predict some universal properties of electron transmission.
They also give very similar expressions for the localization lengths (LL):
$\xi_M\approx(M+1)l$ and $\xi_M\approx[\beta(M-1/2)+1]l$, respectively
($M$ is the number of the propagating channels, $l$ is the phenomenological
mean free path, and $\beta=1 (2)$ for orthogonal (unitary) symmetry class).
In either approach, however, the phenomenological $l$ is viewed as a fixed
parameter. The question how LL explicitly depends not only on
energy $E$, but also on the coupling constants and the type
of boundary conditions in Q1D disordered systems is left out in these
analyses. What we are trying to point out is that in spite of the progress
which has been made to towards a characterization of the localization
in Q1D systems, microscopic analytic studies of LL as a quantum
parameter of fundamental importance, has still not been achieved.

The first step in this direction was done by Dorokhov
in Ref. \cite{dor}, who calculated the LL of $M$ random tight-binding
(TB) chains with random site-energies. The LL in a weak disordered regime was obtained by the author for a Q1D
wire with $M$ channels, is independent of the number of
channels $M$. This result was questioned by Heinrichs \cite{hein02}, where it was shown that for weak disorder and for coupled
two- and three-chain systems ($M = 2, 3$) the inverse
LL is proportional to $M$, in contrast to the result of Ref.
\cite{dor}. However, this approach, adopted in \cite{hein02} and based on a scattering matrix treatment of conductance,
does not allow author to extend his studies of LL
to Q1D systems with larger numbers of scattering channels
$M$. Recently, progress has
been made in taking into account an arbitrary number of channels in
the calculation of LL. R{\"o}mer and Schulz-Baldes \cite{shultz}
using a perturbative formula for the lowest Lyapunov exponent (the inverse LL)
for Q1D TB Anderson strip model with PB conditions obtained LL's dependence on
energy $E$, propagating modes $M$ ($M$ is even) and disorder strength $W$. A \textit{non-perturbative} analytical approach, based on the Green's
function formalism to solve the Dyson equation in Q1D and two-dimensional
(2D) disordered
systems without any restriction on the numbers of impurities and modes, was
developed in Refs. \cite{gas08,GS09,GCJ11}. For a TB Hamiltonian with several
modes and on-site disorder  
the electron's scattering matrix elements $T_{nm}$ (hence the wire conductance $G=\sum_{nm}T_{nm}T^*_{nm}$ (in units of $e^2/h$)) were analytically calculated for an arbitrary impurity profile without actually determining
the eigenfunctions. In these papers \cite{gas08,GS09,GCJ11} only
HW conditions were discussed, which correspond to arranging the parallel
equidistant chains on a plane.

We have performed a careful numerical analysis and
have derived the LL in a Q1D system with HW and PB
conditions. In our numerical calculations we used Kubo's
formula for computing conductivity. The numerical results
were compared with the existing analytical expressions
of LL, calculated in Refs. \cite{dor,hein02,shultz,gas08,GS09,GCJ11}. Surprisingly, our
numerical calculations show that none of these expressions
for LL fit the numerical data well. Particularly, LLs
calculated in Refs. \cite{dor,hein02,gas08,GS09} result in an incorrect dependence
on $M$, while LLs calculated \cite{shultz,GCJ11} correctly predicted the
$M$ dependence, but failed to provide the exact magnitude of LL.

The discrepancy between the theory and our numerical results, as we will explain later on, is due to two main factors:
(i) Since it is not easy to calculate the right-hand
term of the equation (7) (see below), theoretical calculations assume that in the weak disordered regime the length $\left\langle \ln G \right\rangle$ can be replaced by length $\ln \left\langle G \right\rangle$
and by expanding to lowest order of the powers of the
disorder and, after averaging over realizations one can
get a closed analytical expression for LL in the Q1D system ($\left\langle ... \right\rangle$ denotes averaging over disorder realizations).
However, because of not self-averaging the conductance
$G$, these lengths do not agree with each other and thus
lead to a different answer for LL. Note that the same type
of problem exists also in 1D disordered system where, in
the weak disordered limit, LLs, obtained numerically and analytically,
differ by a factor of 2 (see e.g., Refs. \cite{abrahmas,mc}).
(ii) As follows from the numerical analysis of the relationship
between the different LLs in Q1D systems (see
Eq. (1)), the right hand-side term is not zero. This is
an essential piece of information, which allows us later on to introduce new LLs for different boundary conditions
in the transverse direction, which fit 
the numerical data very well.

It is worth noting that while in 1D the relationships
between the various lengths is well known 
(e.g., $\left\langle \ln G \right\rangle = 4\ln \left\langle G \right\rangle =
 -\frac{1}{2} \ln \left\langle 1/G \right\rangle$), in Q1D, to the best of our knowledge,
no such calculations have been previously reported.
Our first goal consists in checking numerically what 
relationship exists for different lengths in the Q1D
case. Once this is established, motivated by our doubts
about the correctness of LL's results of Refs. \cite{dor,hein02,shultz,gas08,GS09,GCJ11} and
to overcome the difficulty of the discrepancy, we have reconsidered
the calculation of LL for the Q1D TB anisotropic
Anderson model, using the Green's function approach,
developed in Refs. \cite{gas08,GS09,GCJ11}. This is our second and main
goal. The analytical results for LL with HW and PB conditions,
Eqs. (9) and (11), are then compared with numerical
results. Excellent agreement with analytical calculations
can be achieved if one multiplies the expressions
(9) and (11) by a factor 2 and shifts them up by $\xi_1$ for HW and by $\xi_1/2$
for PB conditions, respectively. ${\xi_{1}}= {96\sin^2 k_1}/{W^2}$ is the LL in 1D
disordered system, calculated in the weak disordered regime \cite{thou}.

\textit{Relationship between different lengths in Q1D systems}$.-$
First we study the relationship between
$\left\langle \ln G \right\rangle$ and $\ln \left\langle G \right\rangle$ in Q1D
disordered systems. Our numerical calculations show, that these two lengths are
connected through the relation
\begin{equation}
 \left\langle  \ln G  \right\rangle - 4\ln  \left\langle  G  \right\rangle =C. \label{g}
\end{equation}
The constant $C$ is different for HW and PB conditions and for each case is
determined numerically. $C$ tends to zero in the 1D case as expected.
\begin{figure}
\begin{center}
\includegraphics[width=8.0cm]{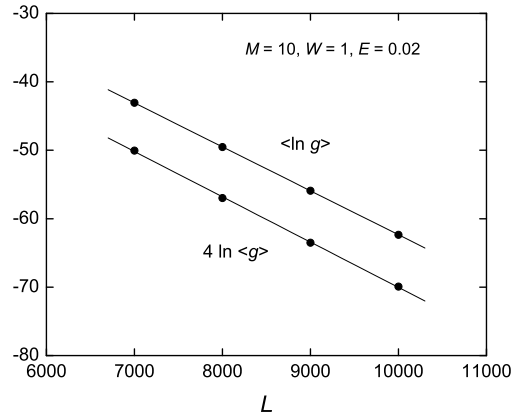}
\caption{Averaged logarithm of the conductance $\left\langle \ln G \right\rangle$ and
logarithm of the average conductance $\ln \left\langle G \right\rangle$ as a function
of the length $L$ of the strip.}
\end{center}
\end{figure}
The length $L$ dependence of $\left\langle \ln G \right\rangle$ and
$4\ln \left\langle G \right\rangle$ is plotted in Fig. 1 for $M=10$ and $W=1$.
The slopes of the two lines are the same within error bars. For each value of $W$
we used $L'$s that ensure that we are well inside the exponential decay
(see Fig. 1).

Some technical details follow:
to obtain the mean values $\left\langle ... \right\rangle$ we have used $10^5$
independent realizations of the disordered strip. Assuming a Gaussian form (as we
have checked) for the logarithm of the conductance, $z \equiv \ln G$, it is
straightforward to show that the average of $G$ is
$\left\langle G \right\rangle = \int_0^M GP(G)dG=\int_0^MP(\ln G)dG$.
Solving the last integral and combining it with the numerical results for
$\left\langle \ln G \right\rangle$ and its variance $\sigma^2$ we get 
\begin{equation}
\left\langle G \right\rangle = \frac{1}{2} \exp ({\left\langle \ln G \right\rangle + \sigma^2/2})
{\rm Erfc} \left[ \frac{\left\langle \ln G \right\rangle + \sigma^2 - \ln M}{\sqrt{2} \sigma} \right],
\end{equation}
where ${\rm Erfc}$ is the complementary error function.

\textit{Q1D TB anisotropic Anderson model}$.-$ Let us discuss a Q1D
disordered lattice of size $L\times M$ described by the standard TB
anisotropic Hamiltonian with nearest-neighbor transfers, $t_{x}$ and $t_{y}$ along the
$x$ and $y$ directions, respectively
\begin{equation}
\begin{split}
{\cal H} &= \sum_{j=1}^{L} \sum_{l=1}^{M} \left| j,l \right\rangle \epsilon_{j,l}
     \left\langle j,l \right| \\
    &- \sum_{j,l} \sum_{\delta = \pm 1} \{ \left| j,l \right\rangle t_{x} \left\langle j +
\delta,l \right| + \left| j,l \right\rangle t_{y} \left\langle j ,l+ \delta \right| \}, \label{ham}
\end{split}
\end{equation}
where $\left| j,l \right\rangle$ is the atomic orbital at site $(j,l)$ and $\epsilon_{j,l}$
is the strength of the random potential at site $(j,l)$, assuming it to be uniformly distributed
in the interval $\left(-\frac{W}{2},\frac{W}{2}\right)$. The disordered region
is connected to perfect leads on both ends, extended to $\pm \infty$ in the $x$ direction.
$L$ is the length of the system and $M$ is the number of modes in the left and right leads.
For simplicity we choose the lattice constant to be equal to $1$. For further calculations we assume the existence of a confining potential $V_c(y_l)$ in the discrete $y$ direction
($y_l=l$, $l=1,2,...,M$) leading to a set of transverse modes, whose actual values depend,
however, on BC conditions. For HW and PB conditions,
the energy of the electron is given by the following dispersion relations
\begin{equation}
E=\left\{ 2t_x \cos k_n+2t_y \cos\frac{\pi n}{M+1} , \; n = 1, 2, ..., M, \; \mbox{ HW}, \atop 
2t_x \cos k_n+2t_y \cos\frac{2\pi n}{M}, \; n = 0, 1, ..., M-1, \; \text{ PB}. \right. \label{0}
\end{equation}
The appropriate eigenfunctions, $\psi_n(y_l)$, of the 1D
Schr{\"o}dinger equation with periodic potential of the chain
of atoms along the $y$ direction with HW and PB conditions
are ($l = 1, 2, ...,M$)
\begin{equation}
\psi_n(y_l)=\left\{ \sqrt\frac{2}{M}\sin \frac{\pi l n}{M+1}\quad \mbox{ HW}, \atop 
\sqrt\frac{2-\delta_{n,0}-\delta_{n,\frac{M}{2}}}{M}\exp{[ i\frac{2\pi nl}{M} ]} \quad \text{ PB}. \right. \label{1}
\end{equation}
Next, closely following Refs. \cite{gas08,GS09,GCJ11}, we evaluate the scattering matrix elements
$T_{nm}$, in the weak disordered regime. The result for the electron transmission amplitude $T_{nm}$ is
\begin{equation}
\begin{split}
T_{nm}& \approx e^{ik_{m}(L-1)}\times \\
&\left\{
\begin{array}{l l}
1-i\frac{\sum^{M}_{l=1}\sum^{L}_{j=1} \epsilon_{j,l}\psi_m(y_l)\psi^*_m(y_l)}{4DL_t\sin k_m}&
\quad \text{if $n=m$},\\ -i\frac{\sum^{M}_{l=1}\sum^{L}_{j=1}\epsilon_{j,l} e^{i\phi_{j}}\psi_n(y_l)
\psi^*_m(y_l)}{2L_t\sqrt{\sin k_n\sin k_m}}& \quad \text{if $n\ne m$},\\
\end{array} \right. \label{nm1}
\end{split}
\end{equation}
where $A_l =\frac{1}{L_t}\sum^M_{n=1}\frac{\sin^2\left({n\pi l}/{L_t}\right)}{\sin k_n}$, $\phi_j=(k_{n}-k_{m})(j-1)$ and $D=1+i\sum^{M}_{l=1}\sum^{L}_{j=1}\epsilon_{j,l}A_l$. The wave numbers $k_{n}$ for the propagating modes are
defined by Eq. (\ref{0}), for HW and PB conditions, respectively.
Similarly, $L_t$ is equal to $M+1$ or $M$ depending on the BC.

The inverse LL $\xi_M$ as a function of the system size $L$
and modes $M$ can be written as
\begin{equation}
\frac{1}{\xi_M}=-\lim_{L\rightarrow\infty}\frac{1}{2ML} \left\langle
\ln \sum^M_{n,m}|T^{(N)}_{nm}|^2 \right\rangle.
\label{xiu1}
\end{equation}
Now, replacing $\left\langle \ln G \right\rangle$ by 
$\ln \left\langle G \right\rangle$ and assuming that for
weak disorder the transmission coefficients are close to 1 and
thus the reflection coefficients are close to zero, we can expand the
right-hand side of Eq. (\ref{xiu1}). Next, after ensemble averaging
over the random potentials $\epsilon_{j,l}$ distributed uniformly
according to the explicit expressions for $T_{nm}$, Eq. (\ref{nm1})
and keeping the terms to order $W^2$, we arrive at the
following expression for the inverse LL
\begin{eqnarray}
\frac{1}{\xi_M}=\frac{W^2}{96M^2}\sum^M_{l=1}\left[
\sum^M_{n=1}\frac{\psi_n(y_l)\psi^*_n(y_l)}{k_n}\right]^2, \label{xi3}
\end{eqnarray}
which is valid for both boundary conditions.

\textit{Hard wall conditions}$.-$ Using the explicit expressions for $ \psi_n(y_l)$
(see Eq. (\ref{1})) and (\ref{xi3}) for the LL $\xi_{M}$ with HW conditions, when
the $M$ channels are propagating, we obtain \cite{GCJ11}
\begin{equation}
\begin{split}
\frac{1}{\xi^{HW}_{M}}&=\frac{W^2}{192M^2(M+1)}\times\\
&\left[
\sum^M_{n=1}\frac{3+\delta_{2n,M+1}}{\sin^2 k_n}+ 2\sum^M_{n<p}\frac{2+
\delta_{n+p,M+1}}{\sin k_n \sin k_p} \right]. \label{xi}
\end{split}
\end{equation}
$k_n$ is the Fermi wave vector of the $n-$th subband (channel) and is determined by
the energy dispersion relation (\ref{0}). For $M=1$ it reduces to the LL $\xi_1$
for a 1D chain. If there is no coupling to the second, third, etc, modes, 
all $k_n$ are equal, and after the summation over the modes, we find from Eq. (\ref{xi})
$\xi^{HW}_M=M\xi_1$.  This result is somewhat expected: it confirms
the prediction of Thouless \cite{thou} that in the limit of weak coupling $\xi^{HW}_M$
must be proportional to $M$. Although one can get two correct limiting values $\xi_1$ and $\xi_{M\gg 1}$ from expression $\xi^{HW}_M$, Eq. (\ref{xi}),
it fails to give
the exact value of LL for an arbitrary $M$.

Our direct numerical computation of the LL for the Anderson model (\ref{ham}) shows that
we can get an almost perfect agreement with the theoretical ${\xi^{HW}_{M}}$, Eq. (\ref{xi})
for $M\ge 2$, if we multiply the latter by a factor $2$ and shift it up by $\xi_1$, i.e.,
redefine new LL $\chi^{HW}_{M}$
\begin{equation}
\chi^{HW}_{M}=\frac{\xi^{HW}_{M}}{2}+\xi_{1}, \quad M\ge 2. \label{xi5b}
\end{equation}
Figure 2 shows the $M$ mode dependence of $\chi^{HW}_{M}$. The solid lines have been computed
from Eq. (\ref{xi5b}) and the dots denote the result of numerical calculations. The good agreement
found fully supports the validity of the analytical expression for $\chi^{HW}_{M}$. The
comparison is free of any adjustable parameter. We have checked that the analytical
expression, Eq. (\ref{xi5b}), agrees very well with the numerical data in the whole
range of vertical hopping parameter $0<t_y < 1$ where the $\chi^{HW}_{M}$ is a linear
function with respect to $M$. The non-linearity starts when $t_y\ge 1$ and for those $t_y$
the numerical and analytical results start behaving differently due to the fact that the validity of formula (\ref{xi5b}) breaks down. For these $t_y \ge 1$ our data show
that $L/\chi^{HW}_{M}\approx 1$, i.e. takes place at the crossover from a Q1D
to 2D system.
\begin{figure}
\begin{center}
\includegraphics[width=8.0cm]{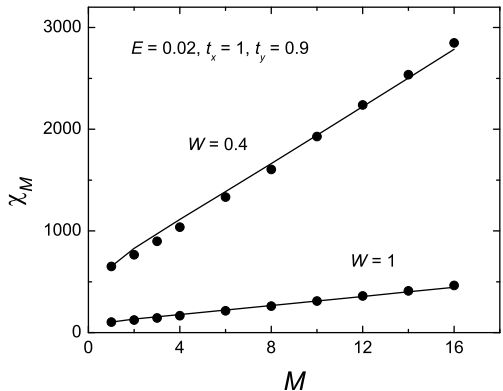}
\caption{The dependence of the localization length $\chi^{HW}_{M}$ on the number of modes
$M$ for disorder W = 0.4 and 1.0. Dots are the numerical results and each data point
corresponds to an average over $10^5$ realizations of disorder. The solid lines represents
the theoretical prediction, Eq. (\ref{xi5a}). At $E=0.02$ all modes are propagating.}
\end{center}
\end{figure}

\textit{Periodic boundary conditions}$.-$ The result for the LL reads
\begin{eqnarray}
\frac{1}{\xi^{PB}_{M}}=\frac{W^2}{96}\left\{
\begin{array}{l l}
\frac{1}{M^3} \left[\sum^{\frac{M}{2}}_{l=0}\frac{2-\delta_{l,0}-
\delta_{l,\frac{M}{2}}}{\sin k_l}\right]^2&, \quad \text{if $M$ even}\\
\frac{1}{(M-1)^3}\left[\sum^{\frac{M-1}{2}}_{l=0}\frac{2-\delta_{l,0}}
{\sin k_l}\right]^2 &, \quad \text{if $M$ odd},\\
\end{array} \right. \label{00}
\end{eqnarray}
where $k_n$ must be defined from the dispersion relation (\ref{0}).

The process of deriving the expression for even $M$ is quite straightforward.
Using the explicit form of the electron wave function $\psi_n(y_l)$, (\ref{1})
and Eq. (\ref{xi3}) yields the desired result. The case for odd $M$ requires
special consideration. First, for the infinitely long periodic system
the conductance $G$ is an odd function of energy, which is
in contrast to the \textit{symmetric} behavior of $G$ with even
$M$ modes. Second, the analysis of the conductance of
the ideal TB model as a function of the energy (at fixed
even $M$) shows that the change from one plateau value
to the next one is $2$ (in units of $e^2/h$), while in the case
of even $M$, it is $1$. Formally this means that $M$ must
be replaced by $(M - 1)$ in the expression of LL with even
number of $M$. This conjecture was numerically tested
and supported by the direct numerical calculation of the
LL (see Fig. 3). It is clear that the difference between
$M$ and $(M - 1)$ is negligible for large $M$, but may not be
negligible for small $M$.

As in the case of HW condition we get an excellent agreement with the theoretical
${\xi^{PB}_{M}}$, Eq. $(\ref{00})$ for $M\ge 2$, if we multiply the latter by
a factor $2$ and shift it up by $\xi_1/2$. The new LL $\chi^{PB}_{M}$ is
\begin{equation}
\chi^{PB}_{M}=\frac{\xi^{PB}_{M}}{2}+\frac{\xi_{1}}{2}, \quad M\ge 2. \label{xi5a}
\end{equation}
In Fig. 3 we have tested the prediction of the analytical theory against the numerical
results where the $M$ mode dependence of $\chi^{PB}_{M}$, Eq. (\ref{xi5a}) is shown.
Solid lines have been computed from Eq. (\ref{xi5a}) and dots denote result of numerical
calculations. The good agreement between simulations (dots) with Eq. (\ref{xi5a}) is evident
for a relatively large range of disorder $W$. In the right panel of Fig. 3 our numerical data for LL
was compared with similar expression
$\frac{1}{\xi}\approx \frac{W^2}{96M^3} \sum^{M-1}_{l,m = 0}\frac{2-\delta_{l,m}}{\sin k_l\sin k_m}$
(dashed line) from Ref. \cite{shultz}. One can see that the slope
of the dashed line agrees with numerical calculations, but certainly there is a problem with accurate
numerical values of LL. To get a correct value for LL for an arbitrary $M$ one needs the dashed line to
shift up by about $0.39\xi_1$. 

In summary, our numerical calculations indicate that in Q1D disordered systems the right-hand side term of Eq. (\ref{g}) is not zero. 
This means that without a vertical shift it is impossible to explain the discrepancy between the theory and numerical data and get the correct analytical expression for LL in the framework of different approaches used in Refs. \cite{dor,hein02,shultz,gas08,GS09,GCJ11}. Based on Eq. (\ref{g}) we have presented the analytical expressions (\ref{xi5b}) and (\ref{xi5a}) for LL which are in excellent agreement with numerical calculations.
%, based on Kubo's formula for conductance.
%We argue that the discrepancy between the theory and numerical data is due to the absence of
%the self-averaging of transmission quantities. This complicates the analytical analysis of the
%disordered systems.

We thank T. Meyer for critical reading of the paper.
The work is supported by FEDER and the Spanish DGI under project no. FIS2010-16430.

\begin{figure}
\begin{center}
\includegraphics[width=10.0cm]{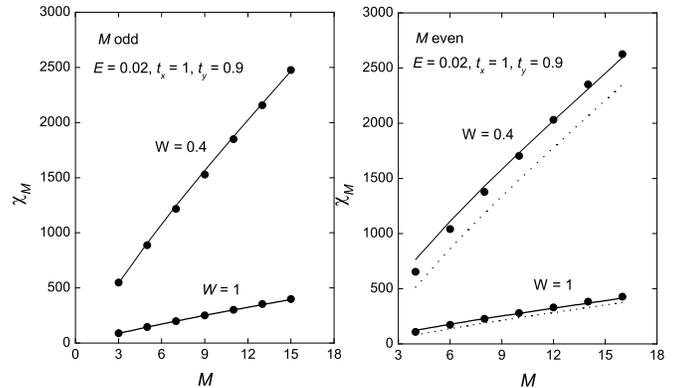}
\caption{$M$ dependence of the localization length defined by Eq. (\ref{xi5a}).
Dots are the numerical results. Left: $M$ is odd. Right: $M$ is even. Dashed line is
presents Eq. (8) of Ref. \cite{shultz}.}
\end{center}
\end{figure}


\begin{thebibliography}{99}
\bibitem{markos} P. Markos, Acta Physica Slovaca {\bf 56}, 561 (2006).
\bibitem{1} P. A. Mello, P. Pereyra, and N. Kumar, Ann. Phys. (NY) {\bf 181}, 290, (1988).
\bibitem{2} C. Beenakker, Rev. Mod. Phys. {\bf 69}, 731 (1997).
\bibitem{dor} O.N. Dorokhov, Phys. Rev. B {\bf 37}, 10526 (1988).
\bibitem{hein02} J. Heinrichs, J. Phys. Cond. Mat. {\bf 15}, 5025 (2003).
\bibitem{shultz} R. A. R{\"o}mer and H. Schulz-Baldes, Europhys. Lett. {\bf 68}, 247 (2004).
\bibitem{gas08} V. Gasparian, Phys. Rev. B {\bf 77}, 113105 (2008).
\bibitem{GS09} V. Gasparian and A. Suzuki, J. Phys. Cond. Mat. {\bf 21}, 405302 (2009).
\bibitem{GCJ11} V. Gasparian, M. Cahay and E. J{\'odar}, J. Phys. Cond. Mat. {\bf 23}, 045301 (2011).
\bibitem{abrahmas} B. S. Andereck and E. Abrahams, J. Phys. C: Solid St. Phys. {\bf 13}, L383 (1980).
\bibitem{mc} A. MacKinnon and B. Kramer, Rep. Prog. Phys. {\bf 56}, 14689 (1993).
\bibitem{thou} D. J. Thouless, Phys. Rev. Lett. {\bf 39}, 1167 (1977).
\end{thebibliography}
\end{document}